# In-Operando magnetometry study on the charge storage mechanism of Sn–Co alloy lithium ion batteries

Qingtao Xia[1,#], Xiangkun Li[1,#], Kai Wang[1,#], Zhaohui Li[1], Hengjun Liu[1], Xia Wang[1], Wanneng Ye[1], Hongsen Li[1], Han Hu[2,*], Jinbo Pang[3], Qinghua Zhang[4], Chen Ge[4], Shandong Li[1,*], Lin Gu[4], Guoxing Miao[5], Shishen Yan[6], and Qiang Li[1,*]

[1]College of Physics, University-Industry Joint Center for Ocean Observation and Broadband Communication, State Key Laboratory of Bio-Fibers and Eco-Textiles Qingdao University, Qingdao 266071, China
[2]State Key Laboratory of Heavy Oil Processing, College of Chemical Engineering, China University of Petroleum (East China), Qingdao 266580, China
[3]Collaborative Innovation Center of Technology and Equipment for Biological Diagnosis and Therapy in Universities of Shandong, Institute for Advanced Interdisciplinary Research (iAIR), University of Jinan, Jinan 250022, China
[4]Beijing National Laboratory for Condensed Matter Physics, Institute of Physics, Chinese Academy of Sciences, Beijing 100190, China
[5]Department of Electrical and Computer Engineering & Institute for Quantum Computing, University of Waterloo, Ontario N2L 3G1, Canada
[6]School of Physics, State Key Laboratory of Crystal Materials, Shandong University, Jinan 250100, China
[#]These authors contributed to this work equally.
Co-corresponding author E-mail addresses: liqiang@qdu.edu.cn (Q. Li), lishd@qdu.edu.cn (S. Li), and hhu@upc.edu.cn (H. Hu).

**Abstract** In view of the long-standing controversy over the reversibility of transition metals in Sn-based alloys as anode for Li-ion batteries, an *in situ* real-time magnetic monitoring method was used to investigate the evolution of Sn-Co intermetallic during the electrochemical cycling. Sn-Co alloy film anodes with different compositions were prepared via magnetron sputtering without using binders and conductive additives. The magnetic responses showed that the Co particles liberated by Li insertion recombine fully with Sn during the delithiation to reform Sn-Co intermetallic into stannum richer phases $Sn_7Co_3$. However, as the Co content increases, it can only recombine partially with Sn into cobalt richer phases $Sn_3Co_7$. The unconverted Co particles may form a dense barrier layer and prevent the full reaction of Li with all the Sn in the anode, leading to lower capacities. These critical results shed light on understanding the reaction mechanism of transition metals, and provide valuable insights toward the design of high-performance Sn alloy based anodes.

**Introduction**

Although today's portable energy storage market has almost been exclusively powered by the rechargeable Li-ion batteries (LIBs)[1-5], the accelerating demands for more advanced electric vehicles and portable electronics continue stimulating extensive research on new electrode materials for breaking the current barriers in energy density and cell durability. Among various anode materials, metallic Sn has been widely considered as a promising anode material for its higher specific capacity than the widely used graphite electrodes[6-10]. However, the pulverization and aggregation induced by the large volumetric variations during the alloying-dealloying processes lead to poor cycling performance. To overcome these issues, inactive metal elements have been introduced to form Sn-M (M=Fe, Co, Ni, and Cu) intermetallic anode materials[11-18], where the inert component acts as a buffer matrix to relieve the volume expansion of tin during cycling. Since Sony proposed the Sn-Co/C battery in 2005[19], various Sn-M materials have been extensively investigated in both academia and industry[10]. However, the mechanism of the reversibility of transition metals in Sn-M intermetallic is still a matter of debate, which restrains the rational design of reliable Sn-M anodes for lithium-ion batteries.

The most widely accepted electrochemical process is a two-step electrochemical reaction, which involves an irreversible initial activation step [Equation 1] followed by the main, reversible, electrochemical process [Equations 2][20-22]

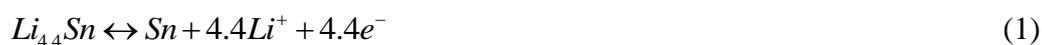

$$Li_{4.4}Sn \leftrightarrow Sn + 4.4Li^+ + 4.4e^- \tag{1}$$

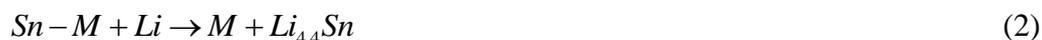

$$Sn-M + Li \rightarrow M + Li_{4.4}Sn \tag{2}$$

By means of *ex-situ* X-ray diffraction (XRD), *ex-situ* Mössbauer spectroscopy and electron paramagnetic resonance spectroscopy, Nwokeke *et al.* claimed that iron nanoparticles are generated during the first discharge of FeSn$_2$ and preserved in the subsequent cycles[23]. However, using *in situ* XRD and *in situ* Mössbauer spectroscopy, Dahn *et al.* found that the Fe nanoparticles formed during discharge can recombine with Sn during delithiation[24, 25]. Interestingly, Whittingham *et al.* revealed that some Fe particles still remain after the first charge by a combination of XRD, X-ray absorption spectroscopy (XAS), and magnetic measurements[26]. As for the Sn-Co alloy, Park *et al.* reported that the CoSn$_2$ shows no recombination during Li extraction through *ex situ* XRD and extended X-ray absorption fine Structure (EXAFS) experiments[15]. In contrast, Han *et al.* showed a complete reversibility of CoSn$_5$ phase by means of *ex-situ* XRD and *ex-situ* XAFS[27]. Additionally, Lee *et al* revealed a partial recombination with Sn during charging in Co$_3$Sn$_2$ alloys by high-resolution transmission electron microscopy (HRTEM) and Co K-edge EXAFS[28]. Obviously, the long-standing controversy over the reversibility of transition metals in Sn-based alloys is caused by the complicated and continuously changing electrochemical processes in the battery, which is beyond the capabilities of many conventional characterization techniques.

Since the magnetism of electrodes containing transition-metals is quite sensitive to the variations of the element valence as well as material structure and morphology during the charge storage processes, magnetometry is a powerful tool for the study of battery processes involving transition-metals[29]. In particular, *operando* magnetic measurements can provide insight into the sequence of redox processes and into the formation of metallic phases[30-34]. Herein, we performed *operando* magnetometry to investigate the reversibility of transition metals in Sn-based alloys operated as anodes. The Sn-Co films with different Co contents were prepared via magnetron sputtering and further assembled into batteries, which are much cleaner compared to the traditional systems with binders and conductive additives therefore more accurate to study the charge storage mechanisms. Depending on their stoichiometry, Sn-Co intermetallic shows different evolution of magnetic signatures. The increased magnetization from the liberated Co particles in $Sn_7Co_3$ film drops to the initial value during the delithiation, demonstrating that Co nanoparticles can fully recombine with Sn in the stannum richer Sn-Co intermetallic. However, the similarly increased magnetization by Li insertion in $Sn_3Co_7$ cannot completely return to the original value, which indicts a partial reversibility of cobalt in the cobalt richer Sn-Co intermetallic. Moreover, the remaining Co particles may form a dense barrier layer, which prevents full reaction of Li with all the Sn in the anode and causes lower capacities. Our results provide a more comprehensive and clearer explanation for the long-standing controversy over the reversibility of transition metals in Sn-based alloys. The better understanding of the charge storage mechanism is important to improve the performance of Sn based alloy anodes.

## Results and Discussion

### Structural and compositional characterization

For a comprehensive understanding of the reaction mechanisms, Sn-Co alloys thin films were prepared by magnetron sputtering, as shown in Figure 1A. Firstly, the self-supported films can exclude the complexity of nanoparticle electrode mixed with binders and conductive additives[35-39]. More importantly, their stoichiometries and particle sizes can be tuned conveniently for control experiments. Since amorphous phases can relax volume change in the case of silicon and tin based anodes[40, 41], we fabricated the films at room temperature to obtain amorphous films. As expected, only the diffraction reflections of the copper foil can be observed in Figure 1B, which indicts that the obtained Sn-Co alloy is nanocrystalline or amorphous. To further confirm their microstructure, selected area electron diffraction (SAED) patterns are shown in insets of Figure 1B. The diffuse rings can be clearly observed, demonstrating the amorphous nature of the deposited films. STEM-energy dispersive X-ray spectroscopy (EDX) elemental mappings in Figure 1 C, D show uniform distribution of Sn and Co elemental signals in the films. The corresponding energy-dispersive spectrum shown in Figure S1 reveals that the Sn–Co alloy film is composed with Sn and Co atomic ratio of 70.2 and 29.8%, respectively (~$Sn_7Co_3$). Figure 1E, F show X-ray photoelectron spectroscopy (XPS) analysis for the chemical composition on the surface of $Sn_7Co_3$ film. The $Sn3d_{5/2}$ peak at 484.7 eV corresponding to metallic tin (Sn (0)) and the $Co2p_{3/2}$ peak at 777.9 eV to metallic cobalt (Co (0)) indicate the formation of Sn-Co alloys. Based on the above characterization results, we can confirm the successful preparation of amorphous Sn$_7$Co$_3$ alloys.

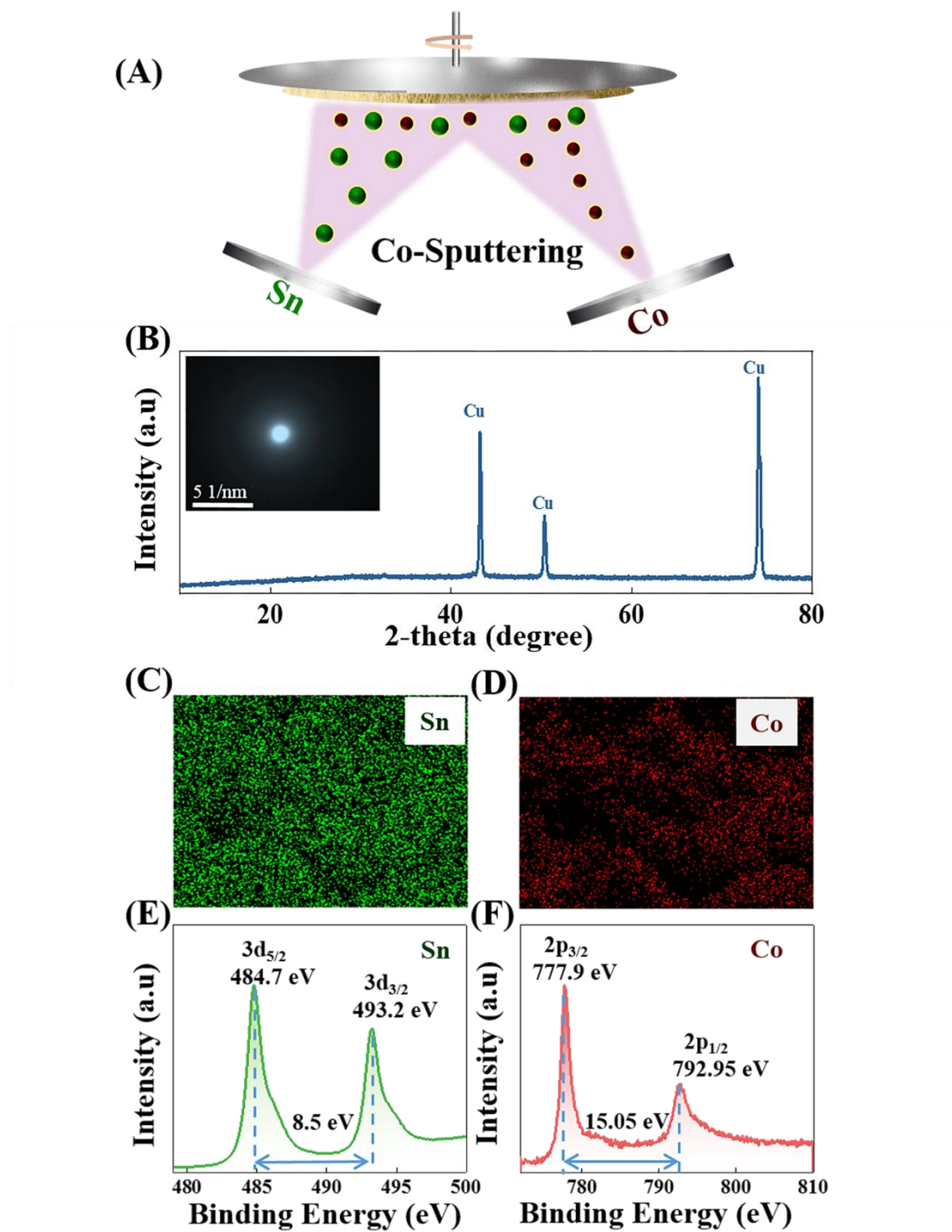

**Figure 1.** (A), Schematic illustration of the fabrication processes for the Sn-Co films. (B), XRD pattern of Sn$_7$Co$_3$ film (the inset is the SAED pattern of Sn$_7$Co$_3$). C, D, EDX mappings of Sn$_7$Co$_3$ pristine sample for the Sn and Co elements, respectively. E, XPS core level Sn 3d spectrum. F, XPS core level Co 2p spectrum of Sn$_7$Co$_3$ pristine sample.

**Electrochemical performance and *Operando* magnetometry**

The electrochemical behavior of the $Sn_7Co_3$ alloy anode was evaluated using cyclic voltammetry (CV) in CR-2032-type coin cells. As displayed in Figure 2A, CV measurements of the first discharge exhibit a very intense peak at ~0.02 V, corresponding to the formation of $Li_xSn$ alloy. In the charge process, the peaks centered at approximate 0.55 and 0.78 V are owing to the extraction of lithium from $Li_xSn$ phases. These results are in line with previous observations for Sn-based negative electrode materials[42]. It is worth mentioning that the discharge peak in the 2nd cycle clearly shifts from 0.02 to 0.2 V, indicating that the restructured alloy (Sn-Co) from the initial cycling would react with Li during following cycling[43]. As correlative results, the plateau regions in the charge-discharge voltage profiles of the $Sn_7Co_3$ anodes in Figure 2B are in good accordance with the current peaks in the CV curves. Additionally, the $Sn_7Co_3$ electrode shows initial discharge and charge capacities of 558 and 365 mA h $g^{-1}$, respectively, corresponding to an initial coulombic efficiency of 65.4%. The irreversible capacity loss in the 1st cycle is mainly caused by the formation of SEI[11].

To investigate the internal structural and chemical evolution during electrochemical cycling, we firstly carried out *ex situ* XRD and XPS measurements of the $Sn_7Co_3$ alloy anode at various states of lithiation as shown in Figure S2 and Figure S3. However, no phase changes of the amorphous alloy film could be detected by XRD. Similarly, there was no change between Co 2p XPS spectra for the pristine, lithiation (0.01V) and delithiation (2.0 V) states. In order to reveal the internal reaction mechanisms, we further conducted *operando* magnetometry studies on $Sn_7Co_3$ electrodes to correlate the electrochemistry with the magnetic changes. Real-time magnetic responses accompanying the galvanostatic charge-discharge tests were recorded as shown in Figure 2C. A low initial magnetic moment agrees with the weakly paramagnetic nature of pristine $Sn_7Co_3$ at room temperature[44]. With the discharge potential decreasing from the open-circuit voltage (OCV) to 0.26V, the saturation magnetization $M_s$ largely remains unchanged, suggesting that Co remains in the alloy. Upon further lithiation, $M_s$ rises sharply until the terminal discharge potential (0.01V). This variation in the magnetization can be ascribed to the conversion from $Sn_7Co_3$ to $Li_xSn$ alloy and cobalt nanoparticles. The magnetization curves after the full discharge to 0.01 V were further performed at 300 K and 5 K, as shown in Figure 3A, which display superparamagnetic behaviour without hysteresis and magnetic saturation even at high fields[45]. The superparamagnetic behaviour can be attributed to the formation of nanosized Co particles, which can also be demonstrated by the high-resolution bright-field scanning transmission electron microscopy (BF-STEM) images (Figure S4A). Meanwhile, Figure S4B shows the results of electron energy loss spectroscopy (EELS) analysis, which can be approximately classified as $Co^0$ fingerprints[46]. During the following charging to 2 V, the magnetization of the liberated Co particles steadily drops to the initial state. In the subsequent cycles, the magnetization rises and falls reversibly. The full recovery in magnetism demonstrates that Co nanoparticles can fully recombine with Sn in $Sn_7Co_3$ upon cycling, which is in agreement with the reported complete reversibility of

CoSn$_5$[27].

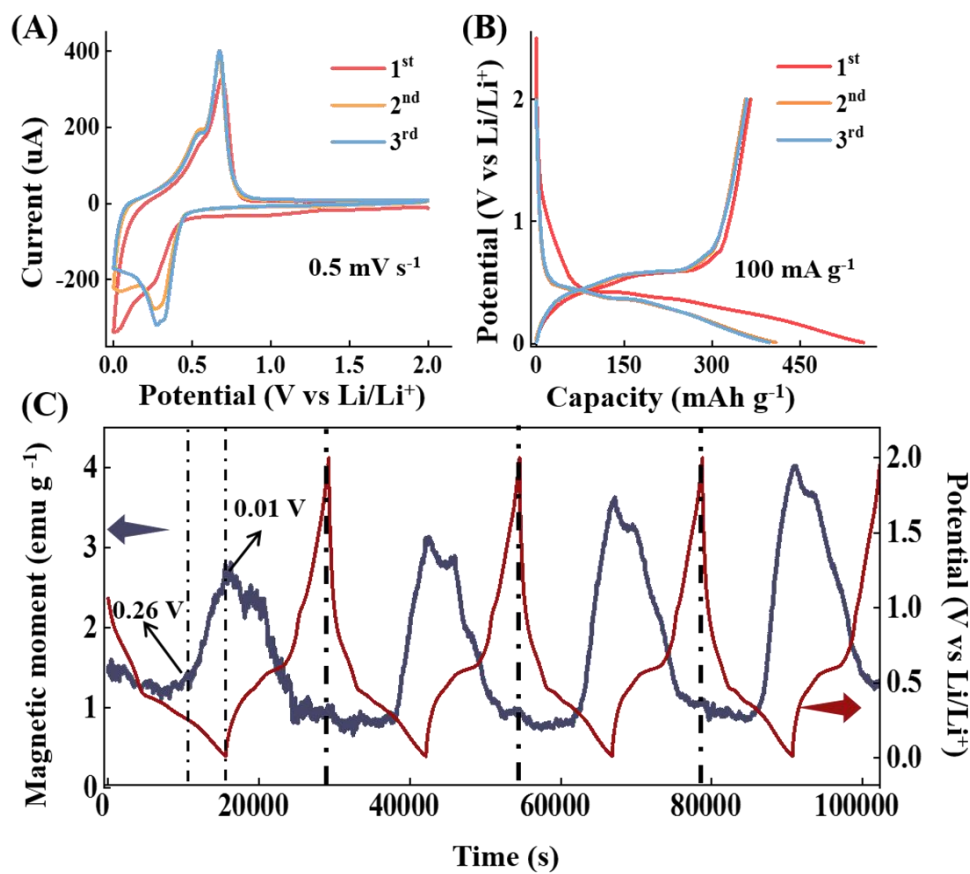

**Figure 2.** A, CV profiles of the Sn$_7$Co$_3$ film with a scan rate of 0.5 mV/s. B, Charge/discharge curves of the first three cycles of the Sn$_7$Co$_3$ film electrode (constant current density: 100 mA g$^{-1}$). C, Variation of the magnetic moment of Sn$_7$Co$_3$ film during galvanostatic cycling under an applied magnetic field of 10000 Oe.

Although the magnetic moment of Sn$_7$Co$_3$ alloy can recover to the initial state by charging up to 2 V, the saturation magnetization of the discharged electrodes gradually increases upon cycling. This increase has also been observed in FeSb$_2$ anode materials, which was attributed to the increase of Fe nanoparticle size[31]. To understand the intriguing increase, we have carried out the zero-field-cooling (ZFC) and field-cooling (FC) measurements on the fully lithiatied material at the end of discharging after n (n=1st, 5th and 10th) cycles. For the ZFC measurement, the sample was first cooled to 5 K in the absence of magnetic fields and the magnetization data were taken upon subsequent heating to 300 K in a finite field (100 Oe). For the FC measurement, the sample was first cooled from 300 K down to 5 K in the presence of 100 Oe magnetic fields, and the data were taken during the heating process in the same field[29, 47]. As shown in Figure 3B, the ZFC/FC curves display typical superparamagnetic characteristics and the ZFC curve exhibits a maximum at the blocking temperature (T$_b$). Interestingly, T$_b$ determined from the ZFC/FC curves increases from 8.7 K to 131 K as the number of cycles increases. From the relationship between the particle size and $T_b$: d= $(150k_B T_b/\pi K_{eff})^{1/3}$, where $k_B$ is the Boltzmann constant and $K_{eff}$ is the effective magnetic anisotropy constant that was assumed to be the same as bulk hcp Co of $4.1\times10^5$ J m$^{-3}$[48, 49] the Co nanoparticles average diameters can be estimated to be 2.4 nm, 3.8 nm, and 5.9 nm using $T_b$ values of 8.7 K, 36 K and 131 K, respectively (Scheme S5). These results indicate that Co particles will gradually become larger upon cycling, which may be related with the volume expansion during the alloying process[31].

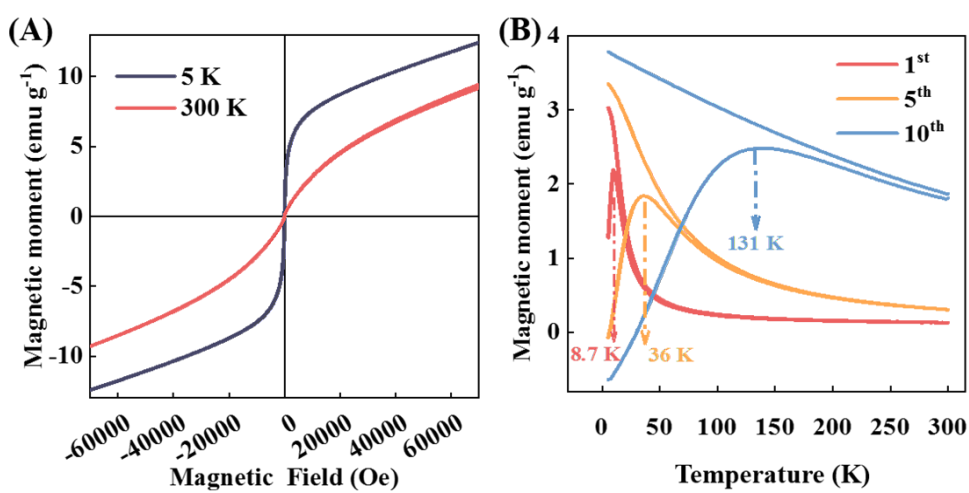

**Figure 3.** A, In-situ magnetic measurements at the end of first lithiation: hysteresis at 300k and 5 K. B, ZFC/FC at 100 Oe on the fully lithiated sample after n (n=1st, 5th and 10th) cycles.

As mentioned in the introduction, some previous work reported that some Sn-Co intermetallic recombines partially irreversibly[15, 28], which may be related to different components. Previous work reported that the cobalt richer phases $Co_3Sn_2$ transformed into $CoSn_2$ during delithiation, preserving a part of the cobalt particles in the subsequent cycles[28]. Therefore, we prepared higher cobalt content anode $Sn_3Co_7$, which was confirmed by EDX shown in Figure S6. *In situ* magnetometry combined with galvanostatic charge-discharge tests were further carried out to study the magnetic evolution in real time (Figure 4A). Figure 4B displays the galvanostatic charge-discharge profiles at current density of 100 mA g$^{-1}$ with a potential range of 0.01-2 V. The initial discharge and charge capacities of 441 and 172 mAh g$^{-1}$ are much lower than the results for $Sn_7Co_3$ due to the higher concentration of cobalt content. At the same time, a higher initial magnetization of 76.6 emu g$^{-1}$ is observed. The magnetization gradually rises to 86.7 emu g$^{-1}$ during the lithium intercalation process due to the precipitation of metallic cobalt, which was confirmed by HRTEM images and SAED pattern (Figure S7A, B). However, it could not completely return to the initial value at the end of the first charging, drastically different from the magnetic evolution of $Sn_7Co_3$. The magnetization of 81.9 emu g$^{-1}$ after the first charge process is much higher than the initial magnetization, indicating that some Co particles are preserved and become inactive in the subsequent cycles. Previous works on Sn-M intermetallic suggested that the extruded inactive transition metals may form a thin layer on the surface of particles or grains and prevent further reaction of the remaining intermetallic material with Li, [25, 42, 50] which leads to a lower capacity as shown in Figure 4B.

To further study the effect of the inactive skin, we compared the electrochemical performance of $Sn_3Co_7$ and $Sn_7Co_3$ with different thickness. By reducing the film thickness, we can reduce the Sn-Co intermetallic grain size, which is expected to weaken the effect of the Co skin on the capacity. Figure 4C, D show the cycling performances of Sn-Co with different thicknesses (50 nm, 100 nm, 200 nm) at 100 mA g$^{-1}$ in the voltage range of 0.01-2.0 V vs. Li/Li$^+$. As expected, the capacity of cobalt richer phases $Sn_3Co_7$ decreases obviously with the increase of thickness (Figure 4C). This result suggests that as the thickness increases, more inner Sn-Co intermetallic are blocked from participating in the reaction with Li due to the dense barrier layer. In contrast, there is almost no variation of capacity in stannum richer $Sn_7Co_3$ with different thicknesses (Figure 4D), which is due to the absence of such barrier layer. These results demonstrate that decreasing the particle and grain size of cobalt richer Sn-Co intermetallic could make more adequate use of the active species.

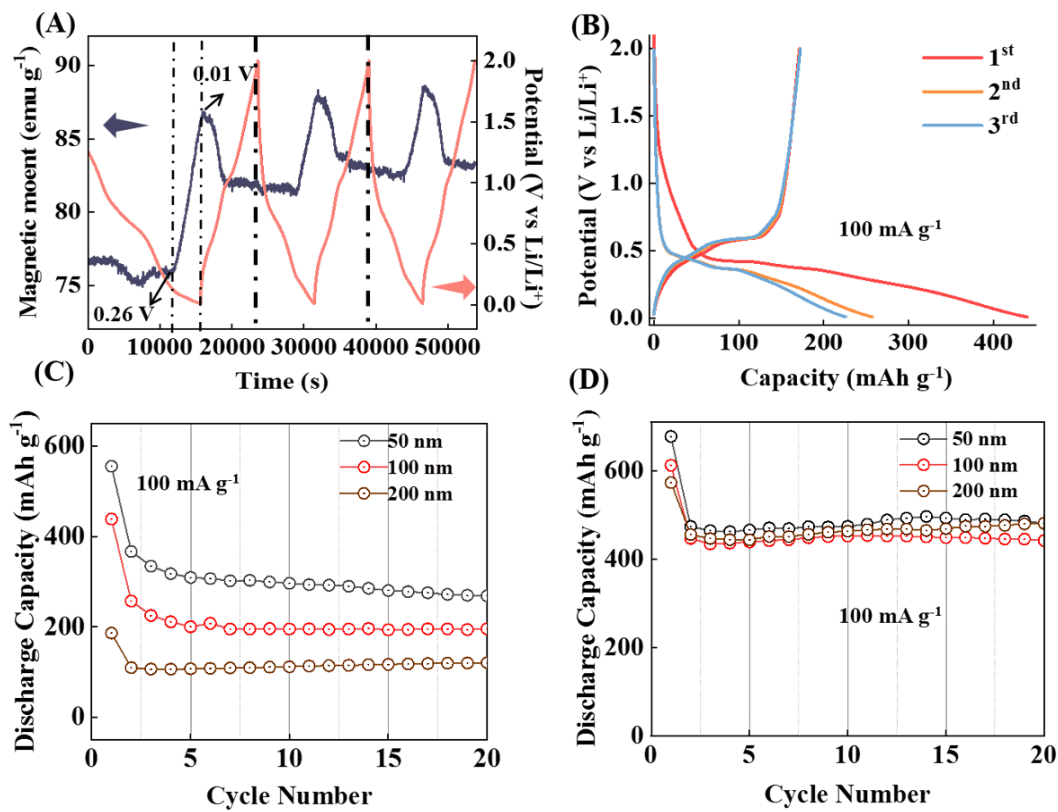

**Figure 4.** A, Variation of the magnetic moment of $Sn_3Co_7$ film during galvanostatic cycling under an applied magnetic field of 10000 Oe. B, Voltage profiles for the first 3 cycles of the $Sn_3Co_7$ film electrode (constant current density: 100 mA g$^{-1}$). C, Discharge capacity of $Sn_3Co_7$ at 50 nm, 100 nm and 200 nm. D, Discharge capacity of $Sn_7Co_3$ at 50 nm, 100 nm and 200 nm.

Base on the above experimental results, the lithiation/delithiation process in Sn-Co intermetallic can be vividly illustrated in Scheme 5A, B. The stannum richer $Sn_7Co_3$ follows the fully reversible electrochemical mechanism:

$$SnCo_y + xLi^+ + xe^- \leftrightarrow Li_xSn + yCo + xe^- \quad (0 < x < 4.4) \tag{3}$$

In contrast, after the cobalt richer phases $Sn_3Co_7$ are transformed into $Li_xSn$ phases and metallic Co nanoparticles, only part of the Co atoms recombine with the released Sn during charging to form Sn-Co alloys. What's more, the excess Co forms a dense barrier layer and prevents further reaction of Li with the remaining intermetallic material. Following the partly reversible electrochemical mechanism:

$$SnCo + xLi^+ + xe^- \rightarrow Li_xSn + Co \quad (0 < x < 4.4) \tag{4}$$

$$Li_xSn + Co \rightarrow (1-y)SnCo + yCo + xLi^+ + xe^- \tag{5}$$

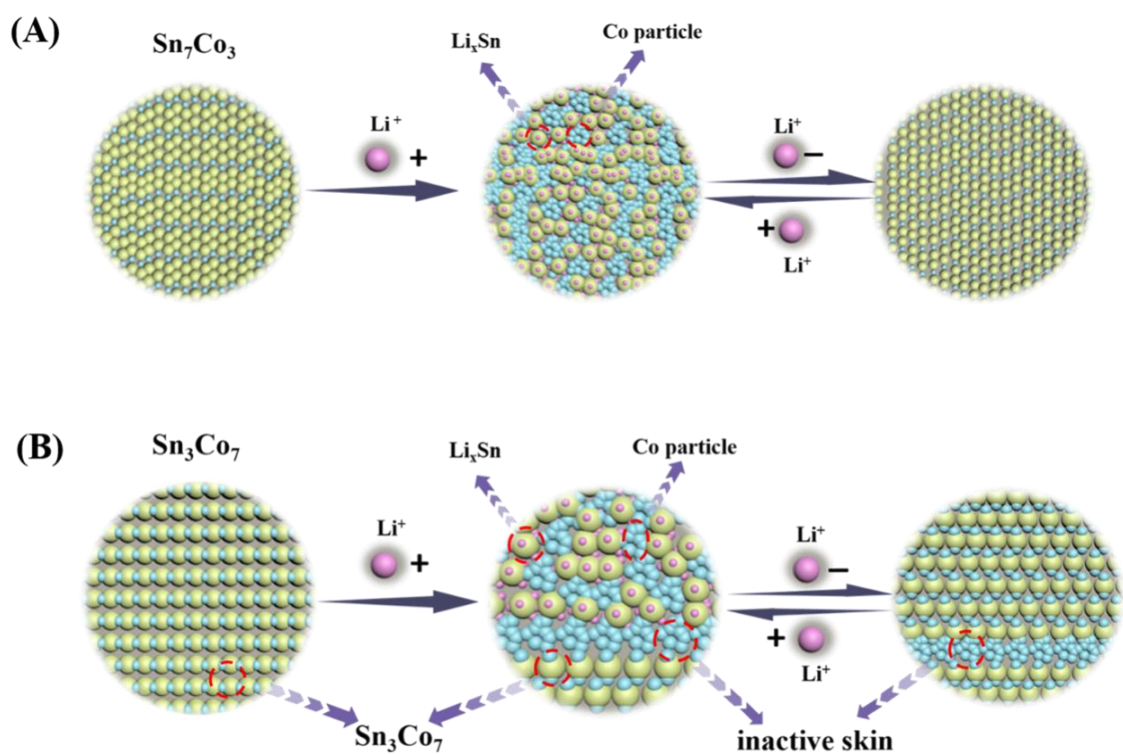

**Scheme 5.** Schematic illustration of the lithiation/delithiation process in A, $Sn_7Co_3$ and B, $Sn_3Co_7$.

We also demonstrated that the Fe particles liberated by Li insertion recombine fully with Sn during the delithiation to reform Sn-Fe intermetallic into stannum richer phases. Similarly, as the Fe content increases, it can only recombine partially with Sn into iron richer phases in iron richer phases (Figure S8). This is very similar to the reaction mechanism of Sn-Co intermetallic. It is worth mentioning that in the Sb-Fe intermetallic, the Fe particles recombine partially with Sb in both antimony richer phases and iron richer phases (Figure S9), which is different from previous work [31]. These differences may be caused by immature operando magnetometry battery assembly process. According to the above results, we can confirm the powerful effect of operating the operando magnetometry to study the reversibility of transition metals in alloys used as anodes.

  **Conclusions**

   In conclusion, to study the electrochemical mechanism of Sn-Co intermetallic, an *operando* magnetometry was used to investigate the magnetic evolution during the electrochemical cycling. Depending on the stoichiometry, Sn-Co intermetallic films deposited by magnetron sputtering ($Sn_7Co_3$ and $Sn_3Co_7$) show different evolution of magnetic signatures. The magnetic responses of stannum richer phases $Sn_7Co_3$ show that the liberated Co nanoparticles can fully recombine with Sn during the delithiation. An increase of Co nanoparticle size during cycling can be revealed by a steady increase of the magnetic moment. As for the cobalt richer $Sn_3Co_7$, the magnetic moment cannot drop to the initial value by charging, indicting a partial reversibility of cobalt. Importantly, the excess Co would form a barrier layer that inhibits the further reaction of Li with the remaining intermetallic material, leading to lower capacities. Decreasing the grain size of intermetallic could weaken the effect of the inactive skin. Our results eliminate the controversies over the charge storage mechanism and provide valuable insights toward the design of high-performance Sn based alloy anodes. These findings also highlight the importance of advanced *operando* magnetometry techniques in the research of energy storage materials related to transition metals.

**Experimental Section**

  *Preparation of Sn-Co alloy thin film electrodes*

The Sn-Co intermetallic compounds ($Sn_7Co_3$ and $Sn_3Co_7$) were obtained by magnetron co-sputtering with a base pressure of $5\times10^{-8}$ Torr at room temperature. Under an Ar gas pressure of 7.5 mTorr, the Sn-Co films were deposited on copper foils by DC sputtering of stannum and cobalt targets (Purity 99.99%). The power applied to the Sn target was 21 W, while that applied to the Co target was 15 W and 30 W for the growth of $Sn_7Co_3$ and $Sn_3Co_7$, respectively. During the co-sputtering process, the substrate was rotated to ensure uniformity of the films.

  *Material characterization*

The crystal structure of the resulted films was characterized by X-ray diffraction (XRD, Bruker D8 advance) with a Cu Kα radiation. The thicknesses of films were measured by an atomic force microscope (AFM; PARK XE7). The loaded mass was measured by the METTLER TOLEDO XP6U with an accuracy of 0.0001 mg. To investigate the morphology and structure of samples, scanning electron microscopy (FESEM, JSM-6700F) and high-resolution transmission electron microscopy (ARM‐200CF, from JEOL) were employed. X-ray photoelectron spectroscopy (XPS) analysis was carried out on a Thermo Scientific ESCALAB 250XI photoelectron spectrometer. To exclude the surface oxidation and the solid electrolyte interphase (SEI), the XPS measurement was performed after 25 nm Ar ion etching. The materials for these *ex situ* characterization were enclosed in airtight bottles under argon atmosphere until measurements.

  *Electrochemical Measurements*

   The electrochemical performances of Sn-Co intermetallic as anode electrode materials were studied by a half cell assembled in an argon-filled glove box. The electrolyte used in this experiment was 1 mol/L LiPF$_6$ in mixture of ethylene carbonate (EC) and dimethyl carbonate (DMC) in the ratio of 1:1 (volume ratio). Electrochemical reactions for the Sn-Co materials were investigated by cyclic voltammetry (CV) at a scan rate of 0.5 mV/s in a voltage window of 0.01−2.0 V versus Li$^+$/Li. The galvanostatic charge/discharge tests were executed on Land battery test system (Land, CT2001A) at various current densities with the cut off voltage range of 0.01–2.0 V. All electrochemical tests were performed at room temperature.

*Magnetic measurement*

The magnetic properties were characterized with a Quantum Design physical property measurement system (PPMS) magnetometer. The cell for *in situ* magnetometry was a pouch-type cell assembled in an argon filled glove box at room temperature. We used Polyethylene terephthalate (PET) sheets to seal the battery and make it flexible for *operando* magnetization measurements. All *in situ* magnetization measurements were taken simultaneously with the electrochemical processes under magnetic fields parallel to the copper foil. The magnetization values given in emu g$^{-1}$ are defined per unit weight of cobalt contained in the Sn-Co intermetallic compounds. The linear magnetic background signals from the other cell assembly components, such as the copper foil, Li metal, and PET sheets, are deducted from the total magnetic moment.

**Conflict of interest**

The authors declare that they have no conflict of interest.

**Acknowledgments**

This work was supported partly by the National Natural Science Foundation of China (11504192, 21975287, 11674186 and 11674187), the National Science Foundation of Shandong Province (ZR2020MA073, ZR2020QE212); Science and Technology Board of Qingdao (16-5-1-2jch). G.X.M. acknowledges Natural Sciences and Engineering Research Council of Canada (NSERC) Discovery grant RGPIN-04178 and the Canada First Research Excellence Fund.

**Author contributions**

Qingtao Xia and Xiangkun Li performed the experiments and wrote the manuscript; Lin Gu and Qinghua Zhang performed STEM measurements; Zhaohui Li, Hengjun Liu, Xia Wang, Kai wang Wanneng Ye, Hongsen Li, Jinbo Pang and Chen Ge contributed to the materials tools and the analysis of the data; Qiang Li, Han Hu and Shangdong Li contributed to the conception of the study; Shishen Yan and Guo-Xing Miao helped perform the analysis with constructive discussions. All authors discussed the results.

**Supporting Information**

I. **Supporting Figures**

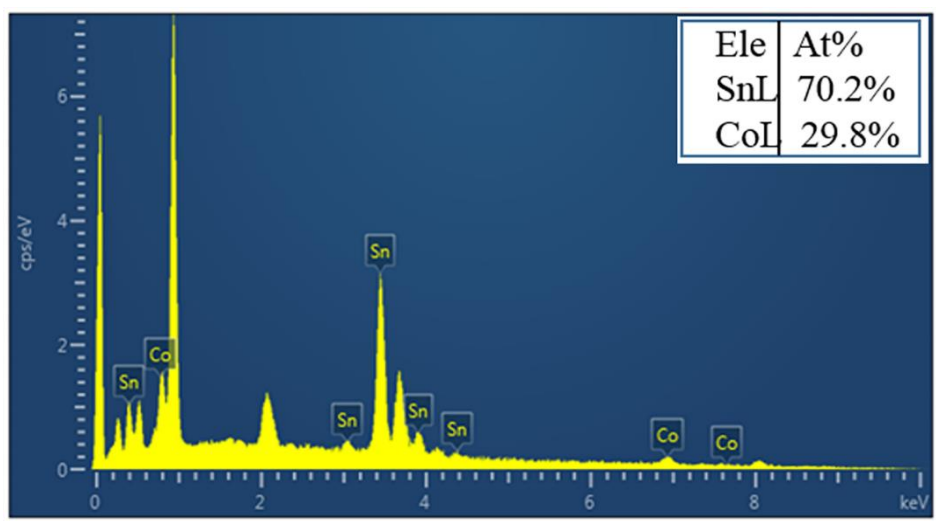

**Figure S1.** EDX analysis of Sn₇Co₃ alloy film electrode.

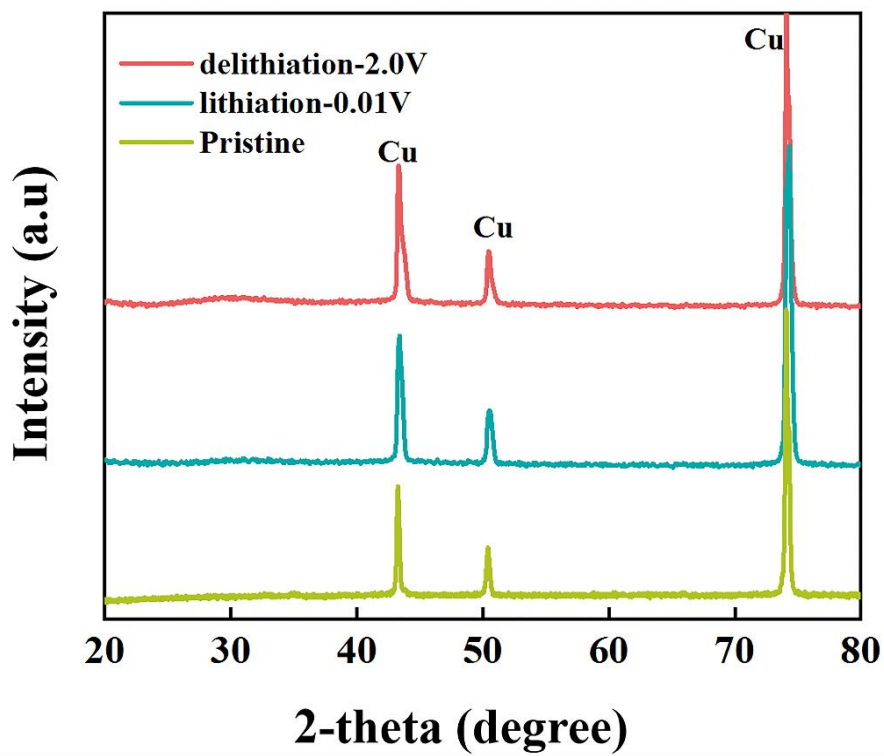

**Figure S2.** *Ex situ* XRD results of the Sn₇Co₃ electrode for the pristine，lithiation (0.01V) and delithiation (2.0V).

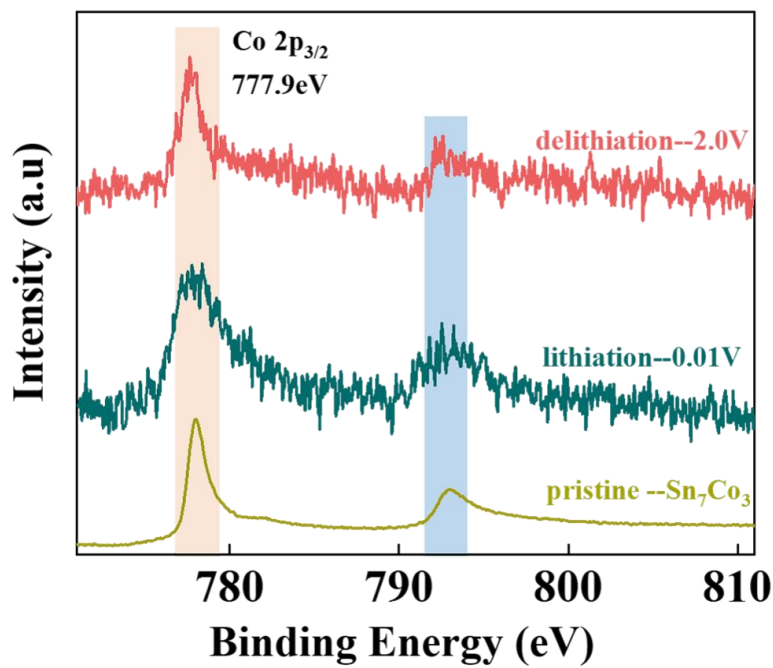

**Figure S3.** *Ex situ* XPS results of the core level Co 2p spectrum for the pristine，lithiation (0.01V) and delithiation (2.0V).

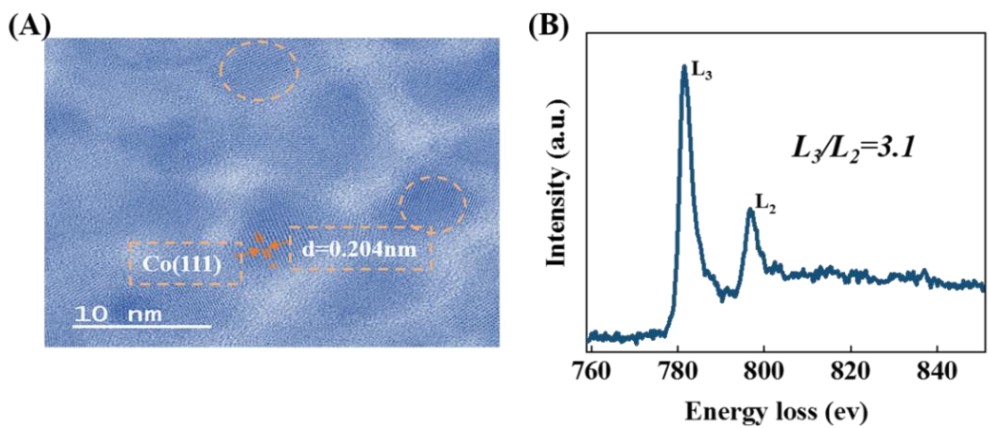

**Figure S4.** (A) High-magnification TEM image of the firstly lithiated $Sn_7Co_3$ electrode. (B) EELS spectra of Co–L2,3 edges recorded from the $Sn_7Co_3$ electrode after the firstly lithiation.

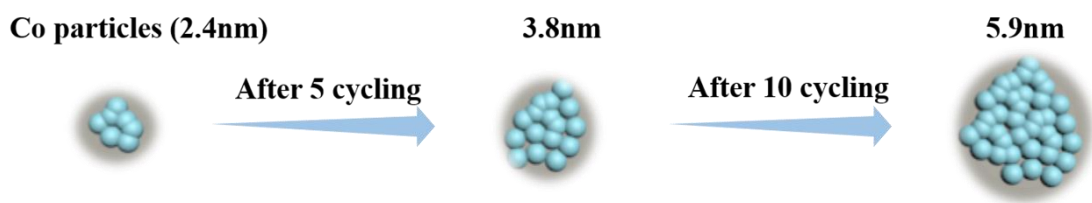

**Scheme S5.** Schematic of the size variation of Co particles during cycling.

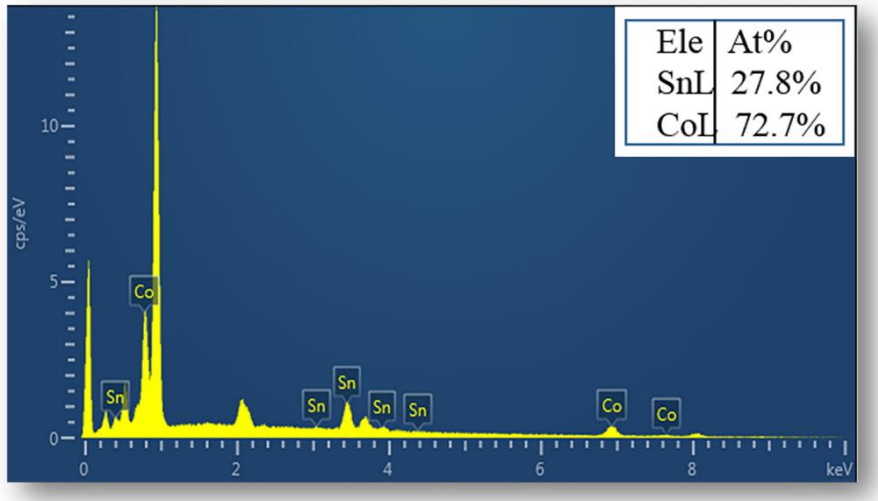

**Figure S6.** EDX analysis of Sn$_3$Co$_7$ alloy film electrode.

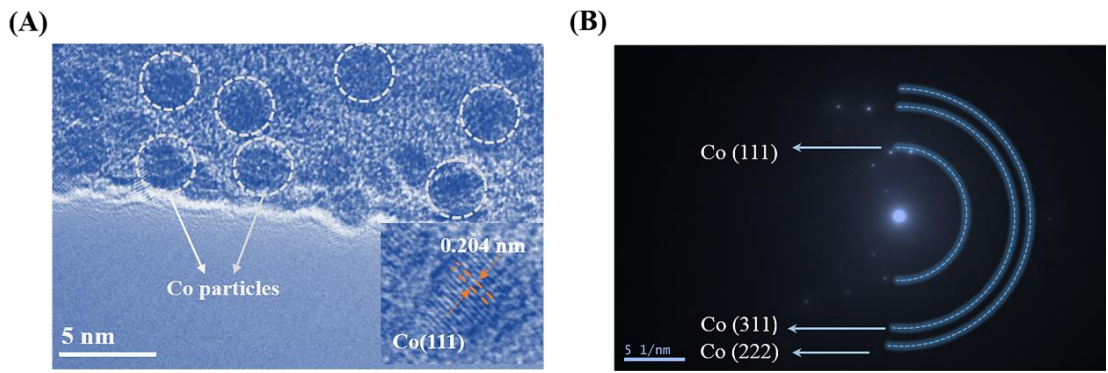

**Figure S7.** (A) High-magnification TEM image of the firstly lithiated 0.01V Sn$_3$Co$_7$ electrode; the inset is a HRTEM image of Co particles. (B) SAED pattern of the lithiated Sn$_3$Co$_7$ electrode.

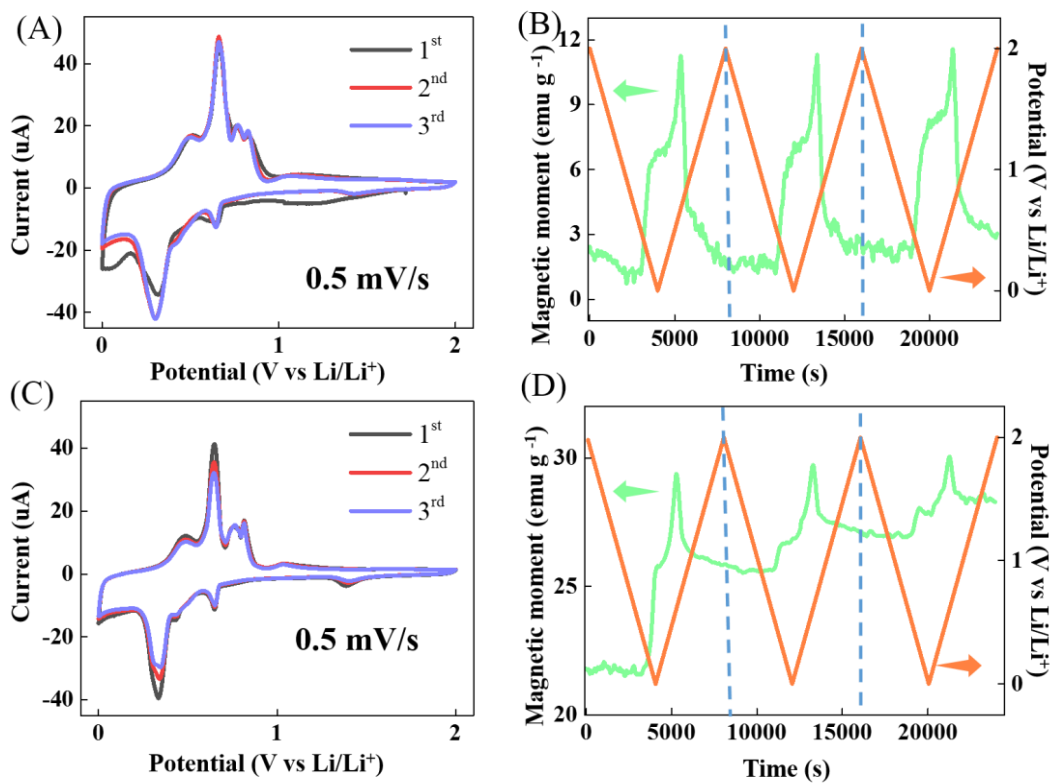

**Figure S8.** (A) CV profiles of the stannum richer Sn-Fe film with a scan rate of 0.5 mV/s (B) Variation of the magnetic moment of stannum richer Sn-Fe film during cyclic voltammetry under an applied magnetic field of 10000 Oe. (C) CV profiles of the iron richer Sn-Fe film with a scan rate of 0.5 mV/s (D) Variation of the magnetic moment of iron richer Sn-Fe film during cyclic voltammetry under an applied magnetic field of 10000 Oe.

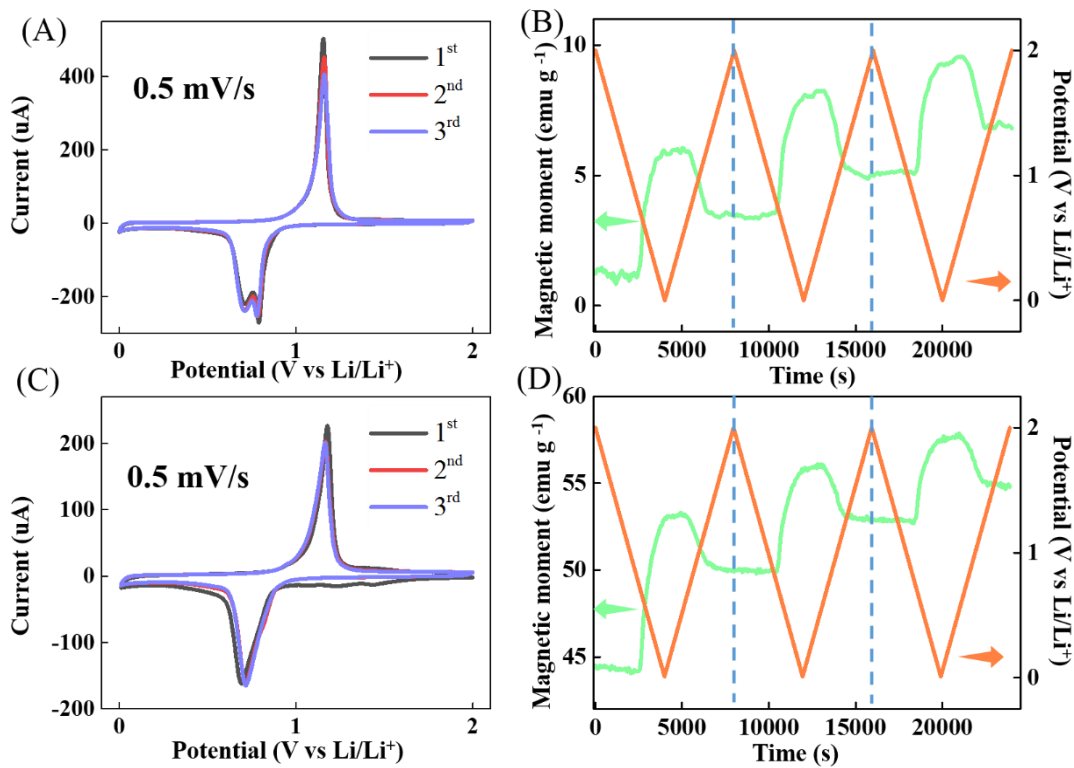

**Figure S9.** (A) CV profiles of the antimony richer Sb-Fe film with a scan rate of 0.5 mV/s (B) Variation of the magnetic moment of antimony richer Sb-Fe film during cyclic voltammetry under an applied magnetic field of 10000 Oe. (C) CV profiles of the iron richer Sb-Fe film with a scan rate of 0.5 mV/s (D) Variation of the magnetic moment of iron richer Sb-Fe film during cyclic voltammetry under an applied magnetic field of 10000 Oe.

Table of content

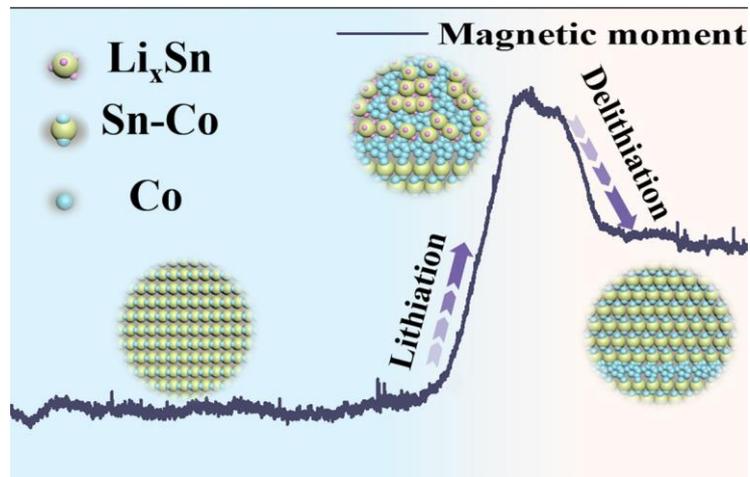

*Operando* magnetometry was used to investigate the charge storage mechanism of Sn-Co intermetallic. The magnetic responses show that the liberated Co particles recombine fully with Sn to reform Sn-Co intermetallic in stannum richer phases. However, it can only recombine partially with Sn in cobalt richer phases. These results eliminate the long-standing controversy over the reversibility of transition metals.